# Weblog Clustering in Multilinear Algebra Perspective

Andri Mirzal

Graduate School of Information Science and Technology,
Hokkaido University, Kita 14 Nishi 9, Kita-Ku, Sapporo 060-0814, Japan

andri@complex.eng.hokudai.ac.jp

## Abstract

This paper describes a clustering method to group the most similar and important weblogs with their descriptive shared words by using a technique from multilinear algebra known as PARAFAC tensor decomposition. The proposed method first creates labeled-link network representation of the weblog datasets, where the nodes are the blogs and the labels are the shared words. Then, 3-way adjacency tensor is extracted from the network and the PARAFAC decomposition is applied to the tensor to get pairs of node lists and label lists with scores attached to each list as the indication of the degree of importance. The clustering is done by sorting the lists in decreasing order and taking the pairs of top ranked blogs and words. Thus, unlike standard co-clustering methods, this method not only groups the similar blogs with their descriptive words but also tends to produce clusters of important blogs and descriptive words.

**Keyword**: clustering method, multilinear algebra, PARAFAC tensor decomposition, weblogs

## I. Introduction

The researches on network clustering have a long tradition in computer science, especially on neighborhood-based network clustering category, where the nodes are being grouped together if they are in the vicinity and have a higher-than-average density of links connecting them [1]. An example of this category is in parallel computing and distributed computation where *n* tasks are divided into several processes that to be carried out by a separate program or thread running on one of *m* different processors [2].

In more general case where the links are weighted according to some particular criteria like similarity measures or distance between two nodes, the clustering tasks can be accomplished by finding good cuts on the network that optimize certain predefined criterion functions. This is usually done by using a technique called *spectral clustering* that has been emerged as one of the most effective tools for document clustering [3]. Under certain conditions, the optimization of the criterion functions in spectral clustering is an equivalent problem to computing the *singular value decomposition* (SVD) of the matrix that captures the relationship between the nodes [5]. But because the vectors produced by SVD are orthogonal, the results usually do not directly correspond to the real clusters and consequently second phase of processing is needed to refine the results. A variety of algorithms (e.g. *k-means*) can be used for this phase [6].

Other famous methods based on similarity matrix can also be used for this category. The direct method is *multidimensional scaling* that simply projects the similarity measures between all node pairs in two-dimensional space [7, 8]. This method is computationally expensive because it has to calculate the similarities of all pairs, thus other more advanced methods that only calculate partial similarities, like k-means [9, 10], *simulated annealing* [11, 10], and *genetic algorithms* [12, 10] are usually being used instead. But due to the incomplete calculations, these methods are subjected to the local optima trap.

In addition to the neighborhood-based network clustering, there is another clustering category that works on labeled-link network; the nodes are in the same group if they share set of similar labels. In online auction networks, this method can be used to find similar users, and then by utilizing user's preferences in buying and selling activities, a recommendation system can be proposed [13]. In hyperlinks environment like web pages, this method can group similar domains with their descriptive hypertexts [14]. Fig. 1 shows the two clustering categories conceptually. In (a), there are two clusters which are well connected within the clusters and only have one link connects them. Conversely, in (b), node 1 and 2 are in the same group due to the similarity in their labels even though they are not connected at all.

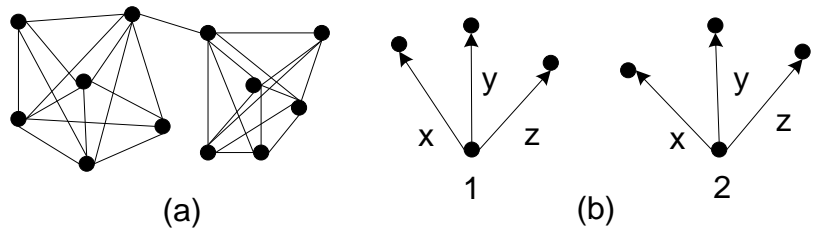

Fig. 1. Clustering based on neighborhood (a), and based on link's labels (b).

In this paper, we describe a clustering method to group the most similar and important weblogs with their descriptive shared words by using PARAFAC tensor decomposition. In the document clustering research, this is known as *co-clustering* problem and has been extensively studied because of its practical uses.

The PARAFAC decomposition is closely related to matrix factorization techniques (see section II.B), but instead of working in two-dimensional document-feature matrix, this method extract three-dimensional document-document-feature tensor from labeled-link network model of the dataset (see Fig. 2). The benefit of this model is the degree of importance of the blogs can be revealed because PARAFAC decomposition gives higher scores to well connected nodes [14, 15]. Thus important blogs tend to be placed in higher ranks.

This work is motivated by Kolda et al. [14] where they model the web pages as the labeled-link network (the nodes are the pages and the labels are the hypertexts), construct the adjacency tensor of the network, and apply tensor decomposition to find the grouping of the pages and the relevant hypertexts. We extend this idea by using contents of the documents instead of the hypertexts, and consequently this task becomes the co-clustering problem. The challenge of weblogs clustering is not a trivial problem. Different from well prepared datasets like TDT2[1] and Reuters[2] document corpora, weblog datasets contain no information about predefined clusters that can be used to compare the results. Thus standard metrics like *F-measure*, *Purity*, *Entropy*, *Mutual Information* [6], and *Accuracy* [16] cannot be used to measure the quality of the results. Also with the tendency of user-centric contents in the Web 2.0 era, the blogs have already become very important information sources. So, the ability to co-cluster the most important blogs will be valuable in assisting the building of the indices for searching.

---

[1] http://www.nist.gov/speech/tests/tdt/tdt98/index.htm
[2] http://kdd.ics.uci.edu/databases/reuters21578/reuters21578.html

## II. Related Works

There are many algorithms or combination of algorithms that can be used in the co-clustering problem. We divide the important algorithms into two categories.

### A. Document-document clustering

Almost all clustering algorithms can be extended in co-clustering problem by finding the most relevant *features* for each cluster. This task is also known as the problem of finding the *cluster centroids*. Note that feature is used instead of term/word to emphasize the fact that the documents actually being characterized by their discriminative features. These features are not limited to words and phrases only, but also formatting tags (e.g. in html files), and even (although very rare) some special characters or punctuations.

There are two classes of algorithms that are widely being used in this category [6]. The first is the *discriminative algorithms* that are based on the similarity measures of document pairs, for example: multidimensional scaling, k-means algorithms, simulated annealing, and genetic algorithms. And the second is the *generative algorithms* that make assumption about the distribution of the data to create model estimation and then use iterative procedure to alternate between the model and the document assignment. The examples are Gaussian model [17], expectation-maximization (EM) algorithm [18], von Mises-Fisher (vMF) model [19], and model-based k-means algorithm [20].

### B. Document-feature clustering

In this category, the co-clustering problem is handled directly. Some of the algorithms are spectral clustering, matrix factorization, and phrase-based model [6].

Spectral clustering builds bipartite graph representation of the documents and then finds good cuts on the graph to optimize predefined objective functions. Because there are two type of nodes (document and feature), the result gives the clusters of similar documents with relevant features. Examples of spectral clustering are divide-and-merge method [21] and fuzzy co-clustering [22].

Matrix factorization is a technique to approximate a matrix ($\mathbf{A}$) by the sum of $K$ matrices ($\mathbf{A}_k$, $k = 1, 2, \ldots, K$) that each is produced by cross product of a pair of vectors. This technique not only reduces data size, but also can increase clustering accuracy because it can reveal the latent relationship between different features that coexist in several documents. If the features are words, this technique can solve synonym problem by indexing not only the words appear in the documents but also other words that are mutually coexistence in other documents. After the matrix is factorized, the clustering results are obtained by finding in which factorization group a document or a feature has the highest score. This group is the cluster label for the corresponding document or word. Two famous matrix factorization techniques are SVD and nonnegative matrix factorization (NMF). While SVD produces orthogonal vectors that can contain negative terms, NMF produces nonnegative vectors that are not necessarily be orthogonal. Because of these characteristics, NMF is better than SVD in finding clusters in the document collection [16], and can be directly utilized as a clustering method. The problem associated with NMF is it depends on initialization; the same data with different initializations will produce different results. Some methods can be used to overcome this drawback, like using spherical k-means to produce vector seeds for NMF [23], or implementing initialization strategies to produce stable results [24].

Phrase-based models try to overcome the weakness in the vector space models by not only encoding the words but also the sequence of the words. Two examples are *suffix tree clustering* [25] that uses shared suffixes from sentences to identify base clusters of the documents, and *document index graph* [26] that represents each word as a vertex in a directed multigraph.

## III. Data Preparation

To conduct the experiment, we use two datasets. The first is from technorati's top 100 blogs[3] by fans and top 100 blogs by authorities downloaded on February 6[th], 2009. The number of non overlapping blogs is 147 (successful downloaded blogs is 140) and the number of words is 9689 (600 words after filtering mechanism is applied). The second dataset is from three sites[4] downloaded on the same date. The number of non overlapping blogs is 155 (successful downloaded blogs is 152) and the number of words is 9099 (592 after filtering mechanism is applied). Note that because tensor decomposition is an expensive method, the number of used words is limited to only about 600.

The filtering mechanism is built to filter out punctuations, tags, stop words, and words that have less discrimination power in clustering like high frequency and very low frequency words, and to stem the words. The stemming algorithm used is porter stemmer [27], a de facto standard stemming algorithm in information retrieval (IR). Further, the filtering is also adapted before the data is fetched by using only contents from blog's rss feeds.

Algorithm 1 is used to fetch feeds from list of the rss feeds in "blogfeedlist" text file, parse the contents, and return a dictionary that holds information about the blog's names as the keys and the contents as the values. A dictionary, also known as associative memories or associative arrays, is a datatype that holds pairs of keys and values. Unlike vector or array which is indexed by integers, a dictionary is indexed its keys. In Algorithm 1, **blogscontent** (a dictionary) holds blog's titles as the keys. The values of **blogscontent** itself are also dictionaries, which hold every word in a blog feed as the keys and the number of the appearance as the values.

Algorithm 1. Function to create a dictionary of blog's name as the keys and their contents as the values.

```
1.    function getFeedsContent(blogfeedlist = readFile("blogfeedlist")) {
2.        blogscontent = dictionary();
3.        for each feedurl in blogfeedlist {
4.            try:
5.                title = getTitle(feedurl);
6.                blogscontent [title] = wordCount(feedurl);
7.            exception:
8.                printscreen("Failed to fetch the feed");
9.                continue to the next feedurl;
10.           }
11.       return blogscontent;
12.   }
```

Algorithm 2 describes the *wordCount*() function (used in Algorithm 1) that takes a feed's url and list of stop words[5] as the inputs and returns a dictionary of unique words in the feed as the keys and the number of appearance of those words as the values. Note that the words are weighted based on

---

their locations in the feed; the description is given highest weight, 10, because this is the place where the authors describe their blog's themes. The next location, the title of the post, is given weight of 3, and each word in the post is given weight of 1. The blog title itself is not used because there are many cases where the blog's title doesn't reflect its contents.

Algorithm 2. Function to list unique words in a blog feed and count their appearances.

```
1.      function wordCount(feedurl, stopwords = readFile("stopwords"),
2.                          int descriptionweight = 10, int posttitleweight = 3,
3.                          int wordweight = 1) {
4.          wordscount = dictionary();
5.          content = parse(feedurl);
6.          description = separateWords(content.description);
7.          if (description != null) {
8.             for each word in description {
9.                if (word in stopwords)
10.                  continue to the next word;
11.               word = porterStemmer(word);
12.               wordscount[word] += descriptionweight;
13.            }
14.         }
15.         posttitle = separateWords(content.posttitle);
16.         if (posttitle != null) {
17.            for each title in posttitle {
18.               if (title in stopwords)
19.                  continue to the next title;
20.               title = porterStemmer(title);
21.               wordscount[title] += posttitleweight;
22.            }
23.         }
24.         summaries = separateWords(content.summaries);
25.         if (summaries != null) {
26.            for each word in summaries {
27.               if (word in stopwords)
28.                  continue to the next word;
29.               word = porterStemmer(word);
30.               wordscount[word] += wordweight;
31.            }
32.         }
33.         return wordscount;
34.     }
```

The first step in reading the feed is to parse the content by using *parse*(), a function that reads feed's xml file and stores the returned values in **content** (see section V for the implementation of this function). This variable has several member variables: **description** holds the description of the blog, **posttitle** holds titles of the post, and **summaries** holds summaries of the post. The second step is to separate words by using *separateWords*() that takes string as the input and returns a list of words in that string. We only consider any non alphanumeric characters as the separators, so words such as C++, Yahoo!, and AT&T, will not be correctly recognized as the complete words. The implementation of this function is not shown here because it is trivial to code it in any scripting languages. The last step is to filter the words out if they are in **stopwords** list and stem them before inputting to **wordscount**.

Algorithm 3. Function to create blog-word characteristic matrix (vector space model).

```
1.      function createCharMatrix(blogscontent = getFeedsContent(),
2.                                  double lower = 0.1,
3.                                  double upper = 0.25) {
4.          blogslist = array();
5.          int i = 0;
6.          for each blog in blogscontent {
7.              blogslist[i] = blog;
8.              i++;
9.          }
10.         count = dictionary();
11.         for each blog in blogslist {
12.             for each word in blogscontent[blog] {
13.                 if (blogscontent[blog][word] > 1)
14.                     count[word] += 1;
15.             }
16.         }
17.         uniquewordslist = array();
18.         int i = 0;
19.         for each blog in blogslist {
20.             for each word in blogscontent[blog] {
21.                 percentage = count[word]/length(blogfeedlist);
22.                 if ((word not in uniquewordslist) and (lower < percentage < upper) ) {
23.                     uniquewordslist[i] = word;
24.                     i++;
25.                 }
26.             }
27.         }
28.         blogsmatrix = matrix();
29.         int i, j = 0;
30.         for blog in blogslist {
31.             for word in uniquewordslist {
32.                 if (word in blogscontent[blog]) {
33.                     blogsmatrix[i][j] = blogscontent[blog][word];
34.                     j++;
35.                 }
36.                 else
37.                     blogsmatrix[i][j] = 0;
38.             }
39.             i++;
40.         }
41.         return blogsmatrix;
42.     }
```

Algorithm 3 takes output from Algorithm 1 and returns blog-word characteristic matrix, **blogsmatrix**. This matrix is the vector space model of the dataset. For list of unique words, we filter out words that appear too often by setting the value of *upper* variable and words that appear only in a few blogs by setting the value of *lower* variable (these values are the percentage of the blogs that contain corresponding words). High frequency words are not really useful because they don't distinguish one blog with others, and low frequency words are too unique so that users almost never use them as the query terms. We set the value of *lower* to 0.1 and the value of *upper* to 0.25. This is not the ideal limits; in [10] the author suggests to use the minimum desired cluster size as the lower limit and 0.4 as the upper limit. We use these values to keep the size of datasets small enough to allow PARAFAC decomposition being applied.

Algorithm 4 is used to transform the output of Algorithm 3 (characteristic matrix) into adjacency tensor. This tensor will be decomposed by using PARAFAC algorithm to produce the co-clustering of blogs and shared words in the next section. Note that colon symbol (:) denotes full range of the given index.

Algorithm 4. Function to transform the characteristic matrix into adjacency tensor.

```
1.      function matrixToTensor(blogsmatrix = createCharMatrix()) {
2.          int K = length(blogsmatrix[0][:]);
3.          int I = length(blogsmatrix[:][0]);
4.          adjTensor = tensor();
5.          for int k = 0, 1, …, K-1 {
6.              for int i = 0, 1, …, I-1 {
7.                  adjTensor [i][:][k] = blogsmatrix[:][k];
8.                  adjTensor [i][i][k] = 0;
9.                  if (blogsmatrix[i][k] == 0)
10.                     adjTensor [i][:][k] = 0;
11.             }
12.         }
13.         temp = adjTensor;
14.         for int k = 0, 1, …, K-1 {
15.             for int j = 0, 1, …, I-1 {
16.                 for int i = 0, 1, …, I-1 {
17.                     if (adjTensor[i][j][k] != 0)
18.                         adjTensor[i][j][k] = adjTensor[i][j][k] + temp[j][i][k];
19.                 }
20.             }
21.         }
22.         return adjTensor;
23.     }
```

The data preparation described in Algorithm 1-4 is equivalent to manipulating the blog dataset into labeled link network. Fig. 2 gives an example of the manipulation process and the extraction of the adjacency tensor from the network. Because the network is constructed from bipartite graph, the result is undirected, thus each frontal slice of the tensor (adjacency matrix for each shared word) is a symmetric matrix.

## IV. PARAFAC Tensor Decomposition

The PARAFAC decomposition is a higher-order analogue technique to the SVD, but the vectors produced by the PARAFAC are not generally orthogonal as the case in the SVD [15]. The PARAFAC decomposition approximates a 3-way tensor by the sum of $R$ rank-1 outer products of vectors $\mathbf{h}_r$, $\mathbf{a}_r$, and $\mathbf{t}_r$ as shown in Fig. 3. Vector $\mathbf{h}_r$ is hub vectors, $\mathbf{a}_r$ is authority vectors, and $\mathbf{t}_r$ is term vectors for each rank $r$. PARAFAC decomposition of tensor $\mathbf{X}$ can be written as [14]:

$$\mathbf{X} \approx \lambda \langle \mathbf{H}, \mathbf{A}, \mathbf{T} \rangle \equiv \sum_{r=1}^{R} \lambda_r \mathbf{h}_r \circ \mathbf{a}_r \circ \mathbf{t}_r \qquad (1)$$

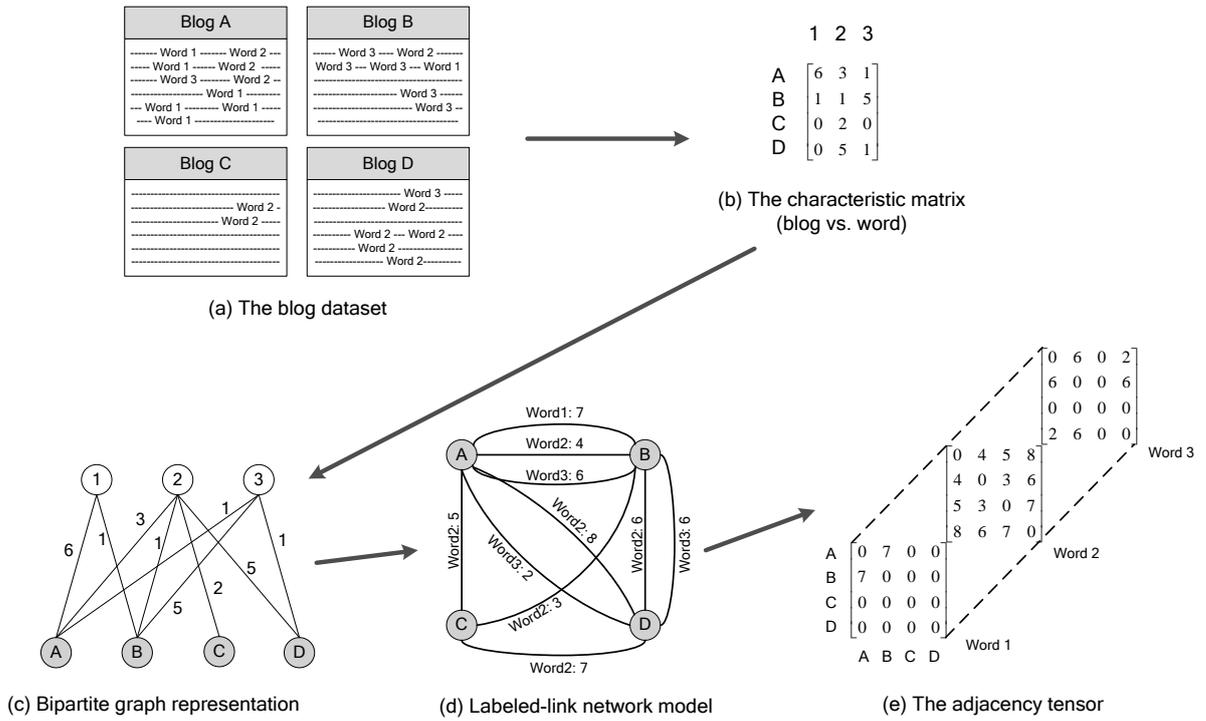

Fig. 2. The labeled-link network construction and adjacency tensor extraction.

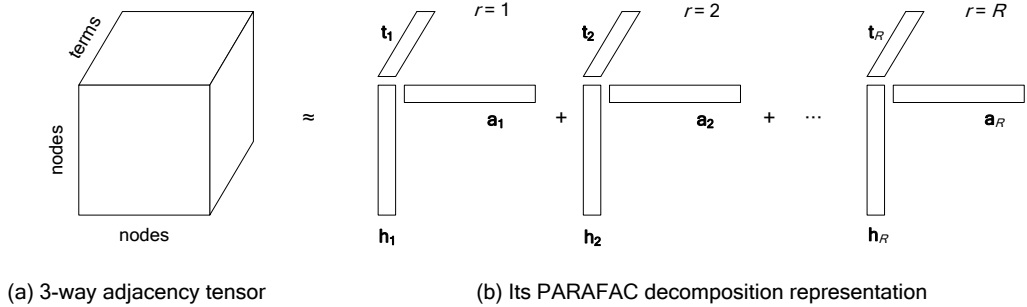

Fig. 3. (a) 3-way adjacency tensor of the network, and (b) its $R$ rank-1 PARAFAC decomposition.

where $\mathbf{H}$, $\mathbf{A}$, and $\mathbf{T}$ is the hub, authority and term matrices of $R$ rank-1 $\mathbf{X}$ decomposition, $\circ$ is outer vectors product, and $\lambda_r$ ($\lambda_1 \geq \lambda_2 \geq \cdots \geq \lambda_R$) is the weight for each group $r$. $\mathbf{H}$, $\mathbf{A}$, and $\mathbf{T}$ are formed by arranging vectors $\mathbf{h}_r$, $\mathbf{a}_r$, and $\mathbf{t}_r$ such that:

$$\mathbf{H} = \begin{bmatrix} \mathbf{h}_1 & \mathbf{h}_2 & \cdots & \mathbf{h}_r \end{bmatrix}, \ \mathbf{A} = \begin{bmatrix} \mathbf{a}_1 & \mathbf{a}_2 & \cdots & \mathbf{a}_r \end{bmatrix}, \text{ and } \mathbf{T} = \begin{bmatrix} \mathbf{t}_1 & \mathbf{t}_2 & \cdots & \mathbf{t}_r \end{bmatrix} \qquad (2)$$

To calculate PARAFAC decomposition, greedy PARAFAC algorithm is used [14]. Algorithm 5 shows the algorithm. Symbol $\|*\|_2$ denotes $L^2$ norm of a vector.

Algorithm 5. Greedy PARAFAC tensor decomposition.

```
1.    function parafac(X = matrixToTensor(), rank, ε) {
2.        int N = length(X[0][:][0]);
3.        int M = length(X[0][0][:]);
4.        double λ =1;
5.        x, y, z, Ψ = array();
6.        H, A, T = matrix();
```

```
7.        for int l = 0, 1, …, rank-1 {
8.          for int n = 0, 1, …, N-1 {
9.              x[n] = 1;
10.             y[n] = 1;
11.          }
12.         for int m = 0, 1, …, M-1
13.             z[m] = 1;
```

14. $\quad\quad\quad \delta = \|\mathbf{x}\|_2 \|\mathbf{y}\|_2 \|\mathbf{z}\|_2 ;$

15. $\quad\quad\quad do \{$

16. $\quad\quad\quad\quad \theta = \delta;$

17. $\quad\quad\quad\quad \mathbf{x} = \mathbf{X} \, \bar{\times}_2 \, \mathbf{y} \, \bar{\times}_3 \, \mathbf{z} - \sum_{r=0}^{l-2} \boldsymbol{\Psi}[r]\mathbf{H}[:][r]\left(\mathbf{y}^T\mathbf{A}[:][r]\right)\left(\mathbf{z}^T\mathbf{T}[:][r]\right);$

18. $\quad\quad\quad\quad \mathbf{y} = \mathbf{X} \, \bar{\times}_1 \, \mathbf{x} \, \bar{\times}_3 \, \mathbf{z} - \sum_{r=0}^{l-2} \boldsymbol{\Psi}[r]\mathbf{A}[:][r]\left(\mathbf{x}^T\mathbf{H}[:][r]\right)\left(\mathbf{z}^T\mathbf{T}[:][r]\right);$

19. $\quad\quad\quad\quad \mathbf{z} = \mathbf{X} \, \bar{\times}_1 \, \mathbf{x} \, \bar{\times}_2 \, \mathbf{y} - \sum_{r=0}^{l-2} \boldsymbol{\Psi}[r]\mathbf{T}[:][r]\left(\mathbf{x}^T\mathbf{H}[:][r]\right)\left(\mathbf{y}^T\mathbf{A}[:][r]\right);$

20. $\quad\quad\quad\quad \mathbf{x} = normalize(\mathbf{x});$

21. $\quad\quad\quad\quad \mathbf{y} = normalize(\mathbf{y});$

22. $\quad\quad\quad\quad \mathbf{z} = normalize(\mathbf{z});$

23. $\quad\quad\quad\quad \delta = \|\mathbf{x}\|_2 \|\mathbf{y}\|_2 \|\mathbf{z}\|_2 ;$

24. $\quad\quad\quad \} \, while \, (|\theta - \delta| < \varepsilon);$

25. $\quad\quad\quad \mathbf{h} = \mathbf{x}; \, \mathbf{a} = \mathbf{y}; \, \mathbf{t} = \mathbf{z}; \, \lambda = \delta;$

26. $\quad\quad\quad \Psi[l] = \lambda;$

```
27.        for int n = 0, 1, ..., N-1 {
28.            H[n][l] = h[l];
29.            A[n][l] = a[l];
30.        }
31.        for int m = 0, 1, ..., M-1
32.            T[m][l] = t[l];
33.      }
34.      return H, A, T, Ψ ;
35.    }
```

There are 3 types of tensor-vector multiplications in Algorithm 5: $\bar{\times}_1$, $\bar{\times}_2$, and $\bar{\times}_3$. A 3-way tensor if multiplied by a vector will become a matrix. But different from a matrix-vector multiplication, the tensor has 3 dimensions, so there are 3 possibilities of multiplying a tensor with a vector. The operation $\bar{\times}_1$, $\bar{\times}_2$, and $\bar{\times}_3$ accommodate these possibilities. Eq. (3) gives examples of $\bar{\times}_1$, $\bar{\times}_2$, and $\bar{\times}_3$ operations, where $\mathbf{X}$ is a 3-way tensor, $\mathbf{u}$ is a vector, and $\mathbf{H}_i$ ($i = 1, 2, 3$) are matrices.

$$\mathbf{H}_1 = \mathbf{X} \, \bar{\times}_1 \, \mathbf{u} \Rightarrow H_1(i_2, i_3) = \sum_{i_1=1}^{I_1} X(i_1, i_2, i_3) u(i_1), \text{ where } 1 \le i_1 \le I_1$$

$$\mathbf{H}_2 = \mathbf{X} \, \bar{\times}_2 \, \mathbf{u} \Rightarrow H_2(i_1, i_3) = \sum_{i_2=1}^{I_2} X(i_1, i_2, i_3) u(i_2), \text{ where } 1 \le i_2 \le I_2 \quad\quad (3)$$

$$\mathbf{H}_3 = \mathbf{X} \, \bar{\times}_3 \, \mathbf{u} \Rightarrow H_3(i_1, i_2) = \sum_{i_3=1}^{I_3} X(i_1, i_2, i_3) u(i_3), \text{ where } 1 \le i_3 \le I_3$$

There are also sequences of tensor-vector multiplications in Algorithm 5. The results of the sequence multiplications (line 17, 18, and 19) are vectors because the tensor is multiplied twice by two vectors. There is an important issue in the sequence multiplication. As shown in eq. (3), the result of the multiplication always reduces the dimension of the tensor, and because the multiplications are performed one by one, there is possibility that the corresponding dimension no longer exists due to the dimensional reduction. Eq. (4) gives an example of such condition.

$$\mathbf{x} = \left( \mathbf{X} \,\bar{\times}_2\, \mathbf{y} \right) \bar{\times}_3 \, \mathbf{z} \tag{4}$$

The result of $\mathbf{X} \,\bar{\times}_2\, \mathbf{y}$ is a matrix, consequently the third dimension no longer exists. Thus $\bar{\times}_3$ operation can no longer be performed. Eq. (5) gives definition of the tensor-vector sequence multiplications to deal with this problem [28].

$$\mathbf{X} \,\bar{\times}_m\, \mathbf{u} \,\bar{\times}_n\, \mathbf{v} \equiv \begin{cases} \left( \mathbf{X} \,\bar{\times}_m\, \mathbf{u} \right) \bar{\times}_n \, \mathbf{v} & \text{if } m > n \\ \left( \mathbf{X} \,\bar{\times}_m\, \mathbf{u} \right) \bar{\times}_{n-1} \, \mathbf{v} & \text{if } m < n \end{cases} \tag{5}$$

And because in the greedy PARAFAC algorithm $m < n$, the second definition is used, $\mathbf{X} \,\bar{\times}_m\, \mathbf{u} \,\bar{\times}_n\, \mathbf{v} \equiv \left( \mathbf{X} \,\bar{\times}_m\, \mathbf{u} \right) \bar{\times}_{n-1} \, \mathbf{v}$.

Algorithm 6 summarizes the process of weblog clustering from the data preparation to tensor decomposition step.

Algorithm 6. Summary of the weblog clustering using PARAFAC tensor decomposition.

1. Store list of blog rss feeds in "blogfeedlist" file
2. Store list of stop words in "stopwords" file
3. Get blogs content:
   **blogscontent** = *getFeedsContent*(*readFile*("blogfeedlist"))
4. Extract blogs characteristic matrix from contents of the blogs:
   **blogsmatrix** = *createCharMatrix*(**blogscontent**)
5. Transform characteristic matrix into adjacency tensor:
   **X** = *matrixToTensor*(**blogsmatrix**)
6. Pick decomposition rank $R$ and a small constant $\varepsilon$ and do decomposition on **X** to get **H**, **A**, **T**, and **Ψ**:
   **H**, **A**, **T**, **Ψ** = *parafac*(**X**, $R$, $\varepsilon$)

## V. Experimental Results

We decompose the tensor of each dataset into 2, 4, …, 14 groups. Our codes for download and data preparation steps are written in python by using *Universal Feed Parser* module[6] for *parse*() function (see Algorithm 2), and for decomposition step are written in MATLAB by using MATLAB Tensor Toolbox[7] [29]. The codes are executed in notebook with Mobile AMD processor 3000+ and 480 MB DDR RAM. The maximum number of groups, 14, was not chosen but the maximum number that our computer can process due to the memory limitation. The computational time increases rapidly as the number of groups increases, with approximately 1.5 hour for first dataset and 1.8 hour for second dataset for 14-group decomposition. Table 1 and 2 show the results for 4-group decomposition.

---

[6] http://www.feedparser.org/
[7] http://www.models.kvl.dk/source/nwaytoolbox/

Table 1. Four-group decomposition for first dataset.

*First Group*

| Blog | Score | Word | Score |
|---|---|---|---|
| Deadspin | 0.2893 | american | 0.2934 |
| Gizmodo | 0.2683 | cases | 0.2647 |
| PW | 0.2283 | government | 0.2124 |
| NewsBusters.org | 0.2235 | facts | 0.2003 |
| Salon, Glenn Greenwald | 0.2206 | official | 0.1774 |
| Consumerist | 0.2019 | countries | 0.1773 |
| Lifehacker | 0.1927 | president | 0.1682 |
| Gawker | 0.1854 | order | 0.1624 |
| Stepcase Lifehack | 0.1778 | administration | 0.1560 |
| Defamer | 0.1769 | claim | 0.1390 |
| ... | ... | ... | ... |

*Second Group*

| Blog | Score | Word | Score |
|---|---|---|---|
| Deadspin | 0.3219 | american | 0.2917 |
| Gizmodo | 0.2791 | cases | 0.2716 |
| PW | 0.2342 | government | 0.2231 |
| Consumerist | 0.2066 | facts | 0.1950 |
| NewsBusters.org | 0.1999 | official | 0.1815 |
| Lifehacker | 0.1976 | countries | 0.1736 |
| Defamer | 0.1886 | administration | 0.1670 |
| Gawker | 0.1875 | order | 0.1619 |
| Bollywood and Cricket | 0.1821 | president | 0.1499 |
| Jalopnik | 0.1730 | court | 0.1387 |
| ... | ... | ... | ... |

*Third Group*

| Blog | Score | Word | Score |
|---|---|---|---|
| PW | 0.44078 | term | 0.07075 |
| The Unofficial Apple Weblog | 0.29312 | feed | 0.06948 |
| Engadget | 0.28701 | roll | 0.02122 |
| Joystiq [Xbox] | 0.27705 | beginning | 0.01338 |
| Joystiq | 0.27639 | low | 0.01299 |
| Luxist | 0.26193 | sales | 0.01166 |
| Gadling | 0.24983 | white | 0.01149 |
| Download Squad | 0.20966 | bit | 0.01138 |
| TV Squad | 0.20943 | apple | 0.01087 |
| Autoblog | 0.20684 | store | 0.01037 |
| ... | ... | ... | ... |

*Fourth Group*

| Blog | Score | Word | Score |
|---|---|---|---|
| Gizmodo | 0.36531 | apple | 0.33161 |
| Lifehacker | 0.29706 | users | 0.31632 |
| The Unofficial Apple Weblog | 0.28522 | search | 0.22253 |
| Mashable! | 0.26911 | iphone | 0.21675 |
| ReadWriteWeb | 0.20867 | application | 0.19533 |
| Stepcase Lifehack | 0.18556 | windows | 0.17675 |
| The Official Google Blog | 0.17158 | store | 0.15905 |
| Download Squad | 0.16017 | mobile | 0.15565 |
| Free Real Traffic to Your Blog | 0.15640 | download | 0.15297 |
| PW | 0.15460 | social | 0.1491 |
| ... | ... | ... | ... |

Table 2. Four-group decomposition for second dataset.

*First Group*

| Blog | Score | Word | Score |
|---|---|---|---|
| Contrary Brin | 0.28917 | suggestion | 0.34724 |
| Writing well is the best revenge | 0.22315 | nations | 0.22647 |
| WOW! Women On Writing Blog | 0.20731 | american | 0.22327 |
| The Writing Show | 0.19113 | obama | 0.16663 |
| Luc Reid | 0.18307 | politics | 0.16641 |
| Six Pixels of Separation | 0.17633 | president | 0.13191 |
| Silliman's Blog | 0.17114 | america | 0.11149 |
| Write Better | 0.16631 | order | 0.09754 |
| Writer Beware Blogs! | 0.16551 | war | 0.09223 |
| Blog Fiction | 0.16277 | effect | 0.08813 |
| ... | ... | ... | ... |

*Second Group*

| Blog | Score | Word | Score |
|---|---|---|---|
| The Writing Show | 0.28019 | character | 0.41977 |
| Writing well is the best revenge | 0.24459 | novel | 0.32174 |
| WOW! Women On Writing Blog | 0.23399 | agent | 0.20159 |
| Luc Reid | 0.22835 | join | 0.16947 |
| Blog Fiction | 0.19351 | author | 0.13254 |
| Kim's Craft Blog | 0.18505 | chapter | 0.10687 |
| Killer Fiction | 0.17404 | contest | 0.10521 |
| Ask Allison | 0.17174 | mysterius | 0.09959 |
| Writer Beware Blogs! | 0.16436 | editor | 0.09833 |
| Spina Bifida Moms | 0.15873 | literary | 0.09772 |
| ... | ... | ... | ... |

*Third Group*

| Blog | Score | Word | Score |
|---|---|---|---|
| PR 2.0 | 0.43184 | twitter | 0.50665 |
| Six Pixels of Separation | 0.29547 | community | 0.38279 |
| Search Engine Guide | 0.18448 | network | 0.27929 |
| Chris Garrett on New Media | 0.17712 | conversion | 0.25382 |
| Marketing Profs Daily Fix | 0.17524 | facebook | 0.22431 |
| Writing well is the best revenge | 0.17013 | brand | 0.19333 |
| The Writing Show | 0.16423 | customer | 0.17262 |
| WOW! Women On Writing Blog | 0.16418 | connection | 0.16602 |
| The Urban Muse | 0.14707 | relation | 0.12927 |
| Write Better | 0.14459 | advertisement | 0.12124 |
| ... | ... | ... | ... |

*Fourth Group*

| Blog | Score | Word | Score |
|---|---|---|---|
| Poets.org | 0.49855 | poetry | 0.69627 |
| Silliman's Blog | 0.31082 | poem | 0.62861 |
| Poets Who Blog | 0.27703 | celebrity | 0.09291 |
| Harriet | 0.24965 | literary | 0.07892 |
| Poetry & Poets in Rags | 0.21138 | nation | 0.07756 |
| Creative Writing Contests | 0.20100 | american | 0.07611 |
| WOW! Women On Writing Blog | 0.19192 | interview | 0.07394 |
| The Writing Show | 0.19076 | contest | 0.07386 |
| Writerswrite.com's Writer's Blog | 0.18135 | award | 0.06302 |
| Mike's Writing Workshop & Newsletter | 0.17831 | collection | 0.04723 |
| ... | ... | ... | ... |

To evaluate clustering accuracy, we compare the tensor decomposition results with NMF results. The reasons of choosing the NMF are: (1) the NMF is analogue to the PARAFAC decomposition in two-dimensional space (while PARAFAC works in higher dimensions), (2) the NMF is preferable than the SVD because it produces nonnegative vectors that are not necessarily be orthogonal, so the vectors are more corresponding to the topics [16], and (3) the NMF produces superior results, especially if NC weighted scheme is used [16].

Eq. (6) gives formulation of the NMF:

$$\mathbf{C} \approx \mathbf{U}\mathbf{V}^T \tag{6}$$

where $\mathbf{C}$ is the $N \times M$ blog-term matrix, $\mathbf{U}$ is nonnegative $N \times R$ blog-factor matrix, and $\mathbf{V}$ is nonnegative $M \times R$ term-factor matrix. The problem is how to find $\mathbf{U}$ and $\mathbf{V}$ that approximate $\mathbf{C}$. The method used here to find $\mathbf{U}$ and $\mathbf{V}$ is based on *multiplicative update rules* [30]. And in order to get the superior results, NC weighted (NMF-NCW) scheme is chosen. In NMF-NCW, normalized version of $\mathbf{C}$ is used instead.

$$\mathbf{C}^* \approx \mathbf{U}\mathbf{V}^T \tag{7}$$

where $\mathbf{C}^* = \mathbf{C}(\mathrm{diag}(\mathbf{C}^T\mathbf{C}\mathbf{e}))^{-1/2}$. To alleviate the local optima trap problem associated with the NMF, 10 trials are performed for each factorization and the best result is picked as the solution. As stated earlier, standard metrics like F-measure, Purity, Entropy, Mutual Information, and Accuracy, cannot be used because no information about predefined clusters is available. Here we define two metrics to assess the quality of the results without the need of the predefined clusters.

The first metric is *similarity measure* that indicates the similarities between the decomposition/factorization results and the standard measures. In search engine researches, similarity measure is used to compare the results returned by certain ranking algorithm to the standard measures. For example, the results of query "barack obama" returned by a search engine is compared to the user votes (standard measures) for the same query to measure the quality of the ranking algorithm implemented by the search engine. We borrow this idea to formulate the metric. But because the results returned by the tensor decomposition are matrices ($\mathbf{H}$, $\mathbf{A}$, and $\mathbf{T}$), they must be converted into blog and word vectors first by using blog and word queries. And because the queries and the groups to be found can be blog or word vectors, there are four possibilities in the query-result relationships as shown in Table 3.

As the standard measures, because user votes for any specific queries are not available, matrix $\mathbf{C}$ is used instead. This choice is intuitive because entries of $\mathbf{C}$ are exact, so it doesn't produce errors or approximate values. Before similar/relevant groups to the queries can be found, the blog's similarity matrix $\mathbf{B}$ and word's similarity matrix $\mathbf{W}$ must be calculated in advanced. Let $N$ be the number of blogs and $M$ be the number of words, $\mathbf{B}$ is $N \times N$ matrix with its entries defined as:

$$\mathbf{B}(i, j) = \cos \angle \big(\mathbf{C}(i,:), \mathbf{C}(j,:)\big), \ 1 \leq i, j \leq N \tag{8}$$

and $\mathbf{W}$ is $M \times M$ matrix with its entries defined as:

$$\mathbf{W}(p, q) = \cos \angle \big(\mathbf{C}(:, p), \mathbf{C}(:, q)\big), \ 1 \leq p, q \leq M \tag{9}$$

Table 3. Query - result relationship

| Query | Similar/relevant group to be found |
|---|---|
| Blogs | Blogs |
| Blogs | Shared words |
| Shared Words | Blogs |
| Shared words | Shared words |

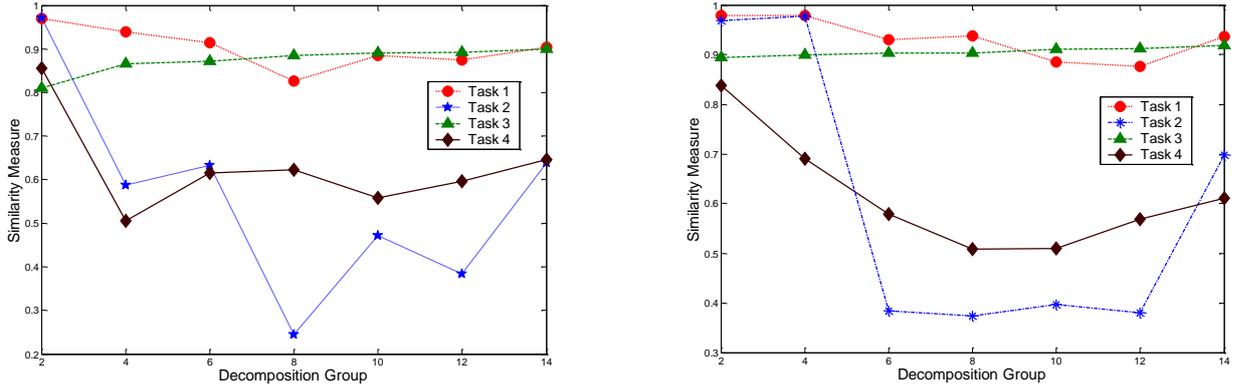

Fig. 4. Similarity measures for PARAFAC decomposition, first dataset (left) and second dataset (right).

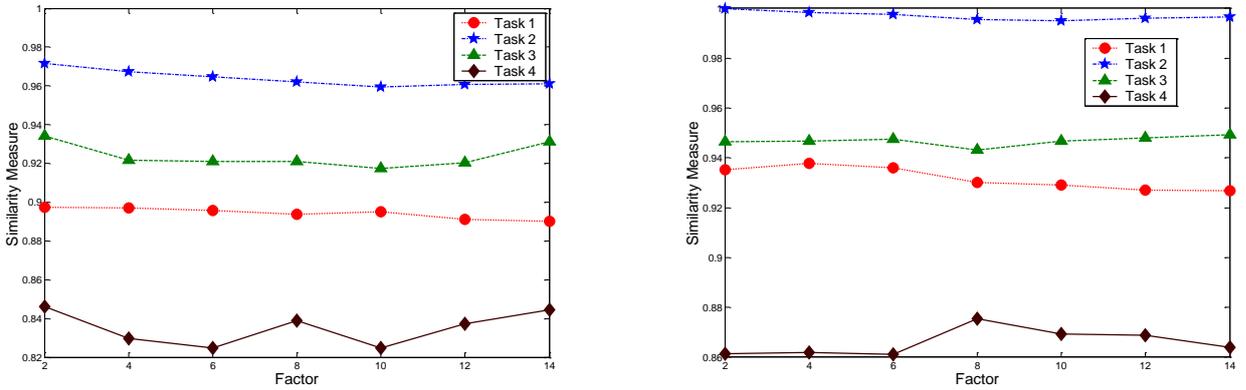

Fig. 5. Similarity measures for NMF method, first dataset (left) and second dataset (right).

Table 4. Standard and decomposition result vectors formulations.

|  | Task 1 | Task 2 | Task 3 | Task 4 |
|---|---|---|---|---|
| $\mathbf{v}_{std}$ | $\mathbf{C}\mathbf{q}_{word}$ | $\mathbf{C}^T\mathbf{q}_{blog}$ | $\mathbf{B}\mathbf{q}_{blog}$ | $\mathbf{W}\mathbf{q}_{word}$ |
| $\mathbf{v}_{dec}$ | $\mathbf{H}^T\mathbf{T}\mathbf{q}_{word}$ | $\mathbf{T}\mathbf{H}^T\mathbf{q}_{blog}$ | $\mathbf{H}\mathbf{H}^T\mathbf{q}_{blog}$ | $\mathbf{T}\mathbf{T}^T\mathbf{q}_{word}$ |

There are two query vectors, $N{\times}1$ blog's query vector $\mathbf{q}_{blog}$, and $M{\times}1$ word's query vector $\mathbf{q}_{word}$, where the entries are ones if the corresponding blogs/words appear in the queries and zero otherwise. Because we are only interested in the average quality of the results, and not in evaluating specific cases, all entries are set to ones: $\mathbf{q}_{blog} = ones(N,1)$ and $\mathbf{q}_{word} = ones(M,1)$. Table 4 gives the formulations of standard vector, $\mathbf{v}_{std}$, and decomposition result vector, $\mathbf{v}_{dec}$, for all four tasks:

1. Task 1: Find the most relevant blogs to words query,
2. Task 2: Find the most relevant words to blogs query,
3. Task 3: Find the most similar blogs to blogs query, and
4. Task 4: Find the most similar words to words query.

For the NMF case, $\mathbf{v}_{std}$ and $\mathbf{v}_{dec}$ formulations are equivalent; simply by replacing $\mathbf{H}$ with $\mathbf{U}$ and $\mathbf{T}$ with $\mathbf{V}$.

The similarities between $\mathbf{v}_{std}$ and $\mathbf{v}_{dec}$ are calculated by using cosine criterion. Fig. 4 shows the results for the PARAFAC decomposition and Fig. 5 for the NMF method.

In the PARAFAC decomposition, there are strong patterns for both datasets; while task 1 and 3 maintain good results for all decomposition groups, task 2 and 4 give unsatisfactory results. Because task 1 and task 3 are the standard way in utilizing blog search engines and apparently there is no practical use of task 2 and 4, these results are promising for indexing purpose.

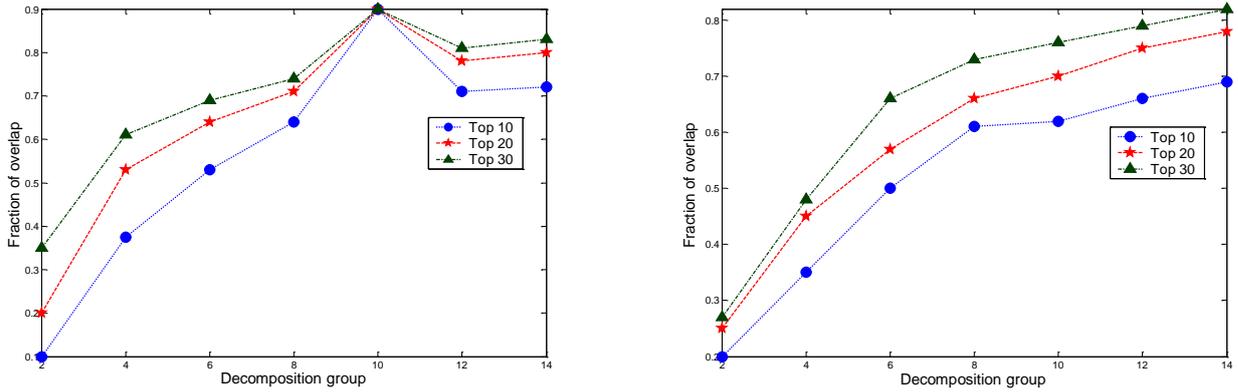

Fig. 6. Fractions of overlap for PARAFAC decomposition, first dataset (left) and second dataset (right).

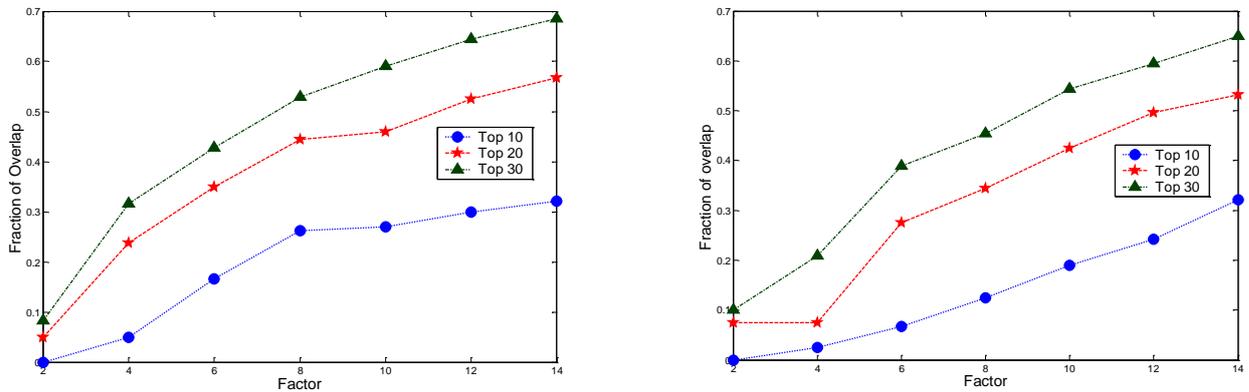

Fig. 7. Fractions of overlap for NMF method, first dataset (left) and second dataset (right).

In general, the NMF produces better and stable results for all tasks in average. But for the important tasks, task 1 and task 3, there are only small differences between these two methods, so it cannot be implied that the NMF is better than the PARAFAC decomposition in the similarity measure metric.

The second metric is *fraction of overlap* that calculates the percentages of overlapping blogs in all blog lists for each decomposition/factorization group. Because the co-clustering is done by sorting the lists of blogs and words in decreasing order (see Table 1 and 2) and the members of each blog list differ only in the orders, if the complete lists are used, the fractions of overlap become 1s. Also because the important aspect of this arrangement is to group the most similar and important blogs with their descriptive shared words, it is natural to use only top ranked blogs. Here we calculate the fractions of overlap for top 10, 20, and 30 blogs only.

In addition to its main function, this metric also has another important role; it can be used to reveal whether a clustering method is able to distinguish the important blogs from the less important ones. If the method has this ability, it will produce high values of overlapping because the important blogs tend to be ranked in the top of the lists. Fig. 6 and 7 show the results for PARAFAC decomposition and NMF method respectively.

In Fig. 6, the overlaps are increasing as the number of decomposition groups and the number of top ranked blogs are increasing. This tendency looks consistent for all plots and datasets. There are anomalies in 10-group in the first dataset. This happens because all 10 blog vectors in the group are repeating each others. If we ignore it, the plots are almost identical for both datasets.

As shown in Fig. 7, the results of the NMF are quite the same. The important difference is the number of overlap produced by the NMF is smaller than the PARAFAC decomposition. There are twofold implications: the NMF produces more distinct topical clusters, and the PARAFAC decomposition produces clusters that contain more information about the degree of importance of the blogs. In the quest of co-clustering the most similar and important blogs with their relevant shared words, the information about the degree of importance is more desirable than the distinct clusters. Thus, the PARAFAC decomposition is preferable than the NMF in this task.

## V. Conclusion

This paper discusses the possibility of using PARAFAC decomposition in co-clustering the most similar and important blogs with their contents. From similarity measure and fraction of overlap calculations, it can be concluded that this method can be used to group the most similar and important blogs with their most descriptive shared words. The main drawback of this method is the computational costs to perform the calculations. We will address this problem in the future research by using optimization techniques, sparse tensor format, and memory management between RAM and harddisk.

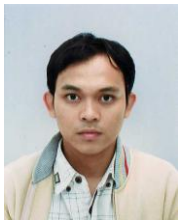

**Andri Mirzal** is a Doctoral course student in Graduate School of Information Science and Technology, Hokkaido University. He received Bachelor of Engineering from Department of Electrical Engineering, Institute of Technology Bandung and Master of Information Science and Technology from Hokkaido University. His research interests are in complex networks, web search engine, and document clustering.